\documentclass[aps,a4paper,superscriptaddress,nofootinbib,twocolumn]{revtex4-1}
\usepackage{amsmath,amssymb,gensymb,graphicx,multirow,dcolumn,bm,latexsym,soul,nicefrac,acronym}
\usepackage[mathscr]{euscript}
\usepackage{appendix}
\usepackage[colorlinks,linkcolor=blue,citecolor=blue,urlcolor=blue]{hyperref}
\usepackage{float}
\usepackage{soul}
\usepackage{verbatim}
\usepackage{color}
\usepackage{acronym}
\usepackage{subfigure}
\usepackage{appendix}
\usepackage{longtable}
\usepackage{footnote}
\usepackage{cancel}
\usepackage{ulem} 

\newcommand{\mnras}{Monthly Notices of the Royal Astronomical Society}

\newcommand{\be}{\begin{equation}}
\newcommand{\ee}{\end{equation}}
\newcommand{\bea}{\begin{eqnarray}}
\newcommand{\eea}{\end{eqnarray}}
\newcommand{\bw}{\begin{widetext}}
\newcommand{\ew}{\end{widetext}}
\newcommand{\nn}{\nonumber}

\newcommand{\eq}[1]{Eq.~(\ref{#1})}
\newcommand{\fig}[1]{Fig.~\ref{#1}}
\newcommand{\tab}[1]{Table.~\ref{#1}}

\newcommand{\HNU}{School of Physics, Henan Normal University, Xinxiang 453007, China}

\newcommand{\TRC}{MOE Key Laboratory of TianQin Mission, TianQin Research Center for Gravitational Physics $\&$  School of Physics and Astronomy, Frontiers Science Center for TianQin, CNSA Research Center for Gravitational Waves, Sun Yat-sen University (Zhuhai Campus), Zhuhai 519082, China}

\begin{document}
\title{Unveiling a multi-component stochastic gravitational-wave background with the TianQin + LISA network}
\author{Zheng-Cheng Liang}
\affiliation{\HNU}
\author{Zhi-Yuan Li}
\affiliation{\TRC}
\author{En-Kun Li}
\affiliation{\TRC}
\author{Jian-dong Zhang}
\affiliation{\TRC}
\author{Yi-Ming Hu}
\email{huyiming@sysu.edu.cn}
\affiliation{\TRC}

\date{\today}

\begin{abstract}
Space-borne detectors, including TianQin and Laser Interferometry Space Antenna (LISA), are tasked with simultaneously observing the Galactic foreground, astrophysical and cosmological stochastic gravitational-wave backgrounds (SGWBs). 
For the first time, we employ a space-borne detector network to identify these SGWBs. 
Specifically, we develop a tailored likelihood for cross-correlation detection with such networks. 
Combined with the likelihood, we use the simulated datasets of the TianQin + LISA network to conduct an analysis for model selection and parameter estimation. 
In our analysis, we adopt an astrophysical background originating from extragalactic white-dwarf binaries, along with a flat cosmological background associated with the early Universe. 
Our results indicate that, after 4 years of operation, the network could detect a single SGWB from either astrophysical or cosmological origins, with an energy density $\Omega_{\rm ast/cos}$ (10 mHz) on the order of $10^{-12}$, despite the presence of a Galactic foreground. 
Furthermore, to distinguish the cosmological background from both a Galactic foreground and an extragalactic background produced by white-dwarf binaries, the energy density $\Omega_{\rm cos}$ should reach around $2\times 10^{-11}$.
\end{abstract}



\maketitle
\acrodef{SGWB}{stochastic \ac{GW} background}
\acrodef{GW}{gravitational-wave}
\acrodef{CBC}{compact binary coalescence}
\acrodef{MBHB}{supermassive black hole binary}
\acrodef{BBH}{binary black hole}
\acrodef{EMRI}{extreme-mass-ratio inspiral}
\acrodef{DWD}{double white dwarf}
\acrodef{GDWD}{Galactic double white dwarf}
\acrodef{EDWD}{extragalactic double white dwarf}
\acrodef{BH}{black hole}
\acrodef{NS}{neutron star}
\acrodef{BNS}{binary neutron star}
\acrodef{LIGO}{Laser interferometry Gravitational-Wave Observatory}
\acrodef{TQ}{TianQin}
\acrodef{LISA}{Laser Interferometry Space Antenna}
\acrodef{TL}{TianQin + LISA}
\acrodef{KAGRA}{Kamioka Gravitational Wave Detector}
\acrodef{ET}{Einstein telescope}
\acrodef{DECIGO}{DECi-hertz interferometry GravitationalWave Observatory}
\acrodef{CE}{Cosmic Explorer}
\acrodef{NANOGrav}{The North American Nanohertz Observatory for Gravitational Waves}
\acrodef{LHS}{left-hand side}
\acrodef{RHS}{right-hand side}
\acrodef{ORF}{overlap reduction function}
\acrodef{ASD}{amplitude spectral density}
\acrodef{PSD}{power spectral density}
\acrodef{CSD}{cross-spectral density}
\acrodef{SNR}{signal-to-noise ratio}
\acrodef{TDI}{time delay interferometry}
\acrodef{PIS}{peak-integrated sensitivity}
\acrodef{PLIS}{power-law integrated sensitivity}
\acrodef{GR}{general relativity}
\acrodef{PBH}{primordial black hole}
\acrodef{SSB}{solar system baryo}
\acrodef{PT}{phase transition}
\acrodef{SM}{Standard Model}
\acrodef{EWPT}{electroweak phase transition}
\acrodef{CPTA}{Chinese Pulsar Timing Arrays}
\acrodef{PTA}{Pulsar Timing Arrays}
\acrodef{RD}{radiation-dominated}
\acrodef{MD}{matter-dominated}
\acrodef{NG}{Nambu-Goto}

\section{Introduction}
The \ac{SGWB} is formed by numerous \ac{GW} sources, each insufficiently strong to be resolvable by detectors. 
The background can be categorized into two main domains: astrophysical- and cosmological-origin. 
The astrophysical \ac{SGWB} is primarily sourced from compact binaries, offering a glimpse into the structure of our galaxy and the history of galaxy mergers~\cite{Cusin:2018rsq,Capurri:2021zli,Bellomo:2021mer}. 
The cosmological \ac{SGWB} stems from early Universe processes, bearing primordial information of the early Universe~\cite{Geller:2018mwu,Jenkins:2018nty,Li:2021iva,Wang:2021djr,Profumo:2023ybp,Schulze:2023ich,Jiang:2024zrb}.

Given the challenge of distinguishing \ac{SGWB} from detector noise, cross-correlation method~\cite{Hellings:1983fr,1987MNRAS.227..933M,Christensen:1992wi,Flanagan:1993ix,Allen:1997ad,Ungarelli:2001xu,LIGOScientific:2017zlf,Hu:2024toa} and null-channel method~\cite{Tinto:2001ii,Hogan:2001jn,Smith:2019wny,Muratore:2021uqj} have been proposed. 
For terrestrial detectors such as ground-based detectors and \acp{PTA}, cross-correlating data from two detectors can distinguish the \ac{SGWB} signal from the detector noise, based on the assumption that the noise in the two detectors is uncorrelated. Ground-based detectors have made strides by setting upper limits on the dimensionless energy parameter $\Omega_{\rm gw}$ within the frequency range of hundreds of Hertz~\cite{LIGOScientific:2009qal,LIGOScientific:2011yag,LIGOScientific:2014gqo,LIGOScientific:2016jlg,LIGOScientific:2017zlf,LIGOScientific:2019vic,KAGRA:2021mth,DeLillo:2023srz}. 
Concurrently, \ac{PTA} collaborations have reported the presence of an \ac{SGWB} in the nano-Hertz ($\rm nHz$) band, marking significant progress in the \ac{SGWB} detection~\cite{NANOGrav:2023gor,Xu:2023wog,EPTA:2023fyk,Reardon:2023gzh}. 
An alternative method can be employed for future space-borne detectors, such as TianQin and the \ac{LISA}, which operate in the milli-Hertz ($\rm mHz$) band. 
Typically arranged in triangular formations, these detectors can establish signal-insensitive data from a null channel. The null channel is capable of monitoring the detector noise. Consequently, the \ac{SGWB} signal can be extracted by auto-correlating the data from the detector~\cite{Adams:2010vc,Adams:2013qma,Caprini:2019pxz,Boileau:2020rpg,Flauger:2020qyi,Banagiri:2021ovv,Cheng:2022vct,Boileau:2022ter,Gowling:2022pzb,Baghi:2023qnq,Wang:2022sti,Hartwig:2023pft,Muratore:2023gxh,Alvey:2023npw,Pozzoli:2023lgz}.

\ac{GW} detectors anticipate encountering a complex blend of \ac{SGWB} originating from diverse sources. 
To extract information regarding the diverse origins from the SGWB, it is essential to conduct component separation~\cite{Parida:2015fma,Sachdev:2020bkk,Biscoveanu:2020gds,Martinovic:2020hru,Sharma:2020btq,Poletti:2021ytu,Kaiser:2022cma,Zhou:2022nmt,Racco:2022bwj,Pan:2023naq,Bellie:2023jlq,Song:2024pnk,Zhong:2024dss,Niu:2024wdi}. 
For space-borne detectors, a predominant foreground is anticipated to arise from unresolved Galactic \acp{DWD}, potentially masking the detection of \acp{SGWB} from other sources~\cite{Bender:1997hs,Nelemans:2001hp,Barack:2004wc,Edlund:2005ye,Ruiter:2007xx,Nelemans:2009hy,Cornish:2017vip,Huang:2020rjf,Liang:2021bde,Staelens:2023xjn,Hofman:2024xar}. 
A single \ac{LISA} can offer the capability to separate backgrounds generated by nearby dwarf galaxies~\cite{Pozzoli:2024wfe} or cosmological sources~\cite{Adams:2013qma,Caprini:2019pxz,Flauger:2020qyi,Boileau:2022ter,Hindmarsh:2024ttn} from the Galactic foreground. Additionally, it can promote further discrimination between astrophysical and cosmological origins~\cite{Boileau:2021sni}.
Nevertheless, the null-channel method employed in single detectors is susceptible to uncertainties in detector noise characterization. 
Such uncertainties can compromise the precision of constraining stochastic backgrounds, potentially leading to a degradation in accuracy by 1-2 orders of magnitude~\cite{Muratore:2023gxh}. 
The cross-correlation method, which harnesses the synergy of multiple space-borne detectors, is crucial to overcome these limitations.

In this paper, we utilize the cross-correlation between space-borne detectors to separate the components of \ac{SGWB}, adopting a Bayesian strategy that is implemented by a  sampling algorithm~\cite{Skilling:2006gxv,Higson:2018cqj,Speagle:2019ivv,Ashton:2022grj}. 
Recognizing the concurrent operational periods and shared detection frequency bands of TianQin~\cite{Li:2024rnk} and LISA~\cite{Colpi:2024xhw}, we leverage the TianQin + LISA (TL) network. 
Accounting for the response and noise budget of TianQin and LISA, we formulate a robust likelihood function, which is designed to handle \acp{SGWB} of arbitrary intensities and incorporates data folding to enhance processing efficiency~\cite{Ain:2015lea,Ain:2018zvo}. 
Additionally, we synthesize detector noise, Galactic foreground, extragalactic background, and cosmological background to generate simulated datasets. 
By applying the likelihood function to conduct Bayesian inference on the simulated datasets, our analysis is twofold: we gauge the network's effectiveness in differentiating between Galactic foreground, astrophysical and cosmological \acp{SGWB} through model selection; and we evaluate the TL network's ability to constrain \ac{SGWB} parameters in the presence of a Galactic foreground via parameter estimation.

The structure of this paper is as follows. 
Sec.~\ref{sec:Key} provides an overview of the essential quantities in \ac{SGWB} detection. 
Building on this foundation, Sec.~\ref{sec:method} introduces the methodology for cross-correlation detection. 
The setup of analysis is laid out in Sec.~\ref{sec:setup}.  
Sec.~\ref{sec:Results} delves into the model selection and parameter estimation for \acp{SGWB}. 
Finally, Sec.~\ref{sec:Summary} presents the conclusion, comparison of studies, and discussion regarding future extensions.

\section{Key quantities for detection}\label{sec:Key}
An \ac{SGWB} originates from a large number of unresolved \ac{GW} events, collectively resembling detector noise\footnote{In many aspects, such as the spectral shape, Gaussianity, stationarity, and anisotropy, the \ac{SGWB} is not noise-like.}. 
This background cannot be fully understood through isolated events, but rather requires statistical analysis. 
One way to characterize \ac{SGWB} is by examining the expectation values of field variables, which includes analyzing the Fourier components of metric perturbations when expanded into plane waves. 
In this section, we will delve into the essential quantities for \ac{SGWB} detection, particularly emphasizing the statistical properties of both the \ac{SGWB} signal and the detector noise.

\subsection{Statistical properties of stochastic background}
In transverse-traceless gauges, the metric perturbations $h(t,\vec{x})$ associated with an \ac{SGWB} can be represented as a superposition of sinusoidal plane waves, which propagate in various directions $\hat{k}$ and at different frequencies $f$~\cite{Misner:1973prb}:
\be
\label{eq:h_ab}
h(t,\vec{x})=\sum_{P}\int_{-\infty}^{\infty}{\rm d}f\int_{S^{2}}{\rm d}\hat{\Omega}_{\hat{k}}\,
\widetilde{h}_{P}(f,\hat{k})\textbf{e}^{P}(\hat{k}) e^{{\rm i}2\pi f[t-\hat{k}\cdot\vec{x}(t)/c]}.
\ee
Here, $\vec{x}$ represents the location where the \ac{GW} measurement is conducted at time $t$, with $c$ signifying the speed of light. 
The spin-2 polarization basis tensor $\textbf{e}^{P}$ corresponds to the two independent polarizations $P=+,\times$ of \acp{GW}. 
The Fourier amplitude $\widetilde{h}_P$ captures the frequency-domain characteristics of \acp{GW} for each given propagation direction.

The statistical properties of an \ac{SGWB} can be characterized by the moments of its Fourier amplitude. 
For the most straightforward case, where the \ac{SGWB} is Gaussian-stationary, unpolarized, and isotropic, the statistical properties can be defined as follows:
\be
\langle\widetilde{h}_{P}(f,\hat{k})\rangle=0,
\ee
\be
\label{eq:Ph}
\langle\widetilde{h}_{P}(f,\hat{k})\widetilde{h}^{*}_{P'}(f',\hat{k}')\rangle
=\frac{1}{16\pi}\delta(f-f')\delta_{PP'}\delta^{2}(\hat{k},\hat{k}')S_{\rm h}(|f|),
\ee
where $\langle...\rangle$ represents the ensemble average, $\delta$ and $\delta_{ij}$ denote the Dirac delta function and Kronecker symbol, respectively. 
To ensure that $S_{\rm h}$ is understood as a one-sided \ac{PSD}:
\be
S_{\rm h}(f)=\lim_{T\to \infty}\frac{1}{T}\left\langle\left|\int_{-T/2}^{T/2}{\rm d}t\,h(t)e^{-{\rm i}2\pi f t}\right|^{2}\right\rangle,
\ee
which involves both polarizations and integrates over the entire sky, a prefactor of $1/16\pi$ has been included.

Alongside the \ac{PSD} $S_{\rm h}$, the dimensionless energy spectrum density $\Omega_{\rm gw}$ is another essential metric, which exhibits a direct relationship with $S_{\rm h}$~\cite{Allen:1996vm,Thrane:2013oya}:
\be
\label{eq:omega_gw}
\Omega_{\rm gw}(f)=\frac{1}{\rho_{\rm c}}\frac{{\rm d}\rho_{\rm gw}}{{\rm d}(\ln{f})}=\frac{2\pi^{2}}{3H_{0}^{2}}f^{3}S_{\rm h}(f).
\ee
Here, $\rho_{\rm gw}$ denotes the \ac{GW} energy density. The critical energy density is defined as $\rho_{\rm c}=3H_{0}^{2}c^{2}/(8\pi G)$, where $G$ is the gravitational constant and $H_{0}$ is the Hubble constant. 

In this paper, we examine the energy spectrum density, which can be categorized into three distinct forms. 
The first one arises from the Galactic foreground. 
The corresponding \ac{PSD} is given by the polynomial expression~\cite{Liang:2024ulf}:
\be
\label{eq:Omega_fg}
S_{\rm h}(f)=
\frac{20}{3}\left[10^{\sum_i a_{i}x^{i}(f)}\right]^{2},
\ee
where the factor of $20/3$ is related to the response of space-borne detectors to \acp{GW} for the all-sky average~\cite{Cornish:2001qi,Huang:2020rjf}, and the function $x(f)=\log(f/10^{-3}\,\,{\rm Hz})$ is defined with polynomial coefficients $a_{i}$. 
Note that the \ac{PSD} of the Galactic foreground is associated with the observation time of detectors, which we will address in Sec.~\ref{sec:components}.
The second one pertains to the astrophysical \ac{SGWB}. 
Its energy spectrum density is characterized by a power-law form:
\be
\label{eq:Omega_ast}
\Omega_{\rm gw}(f)=\Omega_{\rm ast}\left(\frac{f}{f_{\rm ref}}\right)^{\alpha_{\rm ast}},
\ee
where the spectral index $\alpha_{\rm ast}$ can be set to $2/3$ for c binary coalescence within the mHz band~\cite{Phinney:2001di}, and the reference frequency $f_{\rm ref}$ is chosen to be arbitrary. 
Lastly, the cosmological \ac{SGWB} is predicted to have a nearly flat spectrum~\cite{Figueroa:2012kw,Maggiore:2018sht,Auclair:2019wcv}:
\be
\label{eq:Omega_cos}
\Omega_{\rm gw}(f)=\Omega_{\rm cos}.
\ee

\subsection{Signal and noise PSDs}\label{sec:s_n}
Space-borne \ac{GW} detectors, such as TianQin and the \ac{LISA}, are specifically designed to detect \acp{GW} in the $\rm mHz$ band. 
TianQin will feature an equilateral triangle configuration, with three identical satellites orbiting Earth at an arm length of approximately $1.7\times10^{5}\,\, {\rm km}$. 
It will have an orbital period of 3.64 days and will operate on a schedule alternating between ``three months on" and ``three months off"~\cite{TianQin:2015yph}. 
\ac{LISA} also utilizes a triangular configuration, but it will orbit the Sun, trailing behind the Earth by about $20\degree$, 
The three satellites in LISA will maintain a separation of roughly $2\times10^{6}\,\,{\rm km}$ from one another. 
LISA's orbital period is synchronized with that of the Earth, completing one orbit around the Sun within a year. 
It is important to note that while TianQin and LISA can potentially form the TL network for \ac{GW} detection, their joint observation time is limited to only half of their operation time due to TianQin’s working mode. 
This constraint will be generally applied throughout the analysis of this paper.

The \ac{SGWB} signal in a detector channel $I$ can be represented as the convolution of channel response $\mathbb{F}_{I}(t,\vec{x})$ and metric perturbations $h(t,\vec{x})$~\cite{Romano:2016dpx}. 
Given the detector motion, which introduces variations in the response, it is advantageous to focus on a narrow time interval, defined as $[\tau-T/2,\tau+T/2]$, to ensure that the \ac{GW} measurement location and the channel response are nearly static. 
Consequently, the \ac{SGWB} signal can be expressed as:
\bea
\label{eq:ht_sgwb}
\nn
h_{I}(t,\tau)&=&\mathbb{F}_{I}[t,\vec{x}_{I}(\tau)]*h[t,\vec{x}_{I}(\tau)]\\
\nn
&=&
\int_{-\infty}^{\infty}{\rm d}f \, \sum_{P}
\int_{S^{2}}{\rm d}\hat{\Omega}_{\hat{k}}
F_{I}^{P}(f,\hat{k},\tau)\widetilde{h}_{P}(f,\hat{k})\\
&&\times 
e^{{\rm i}2\pi f[t-\hat{k}\cdot\vec{x}_{I}(\tau)/c]},
\eea
where $\tau$ is employed to label the specific time interval, the frequency-domain response $F_{I}^{P}$ is determined by the polarization tensor and the detector channel $I$~\cite{Cornish:2001qi}. 
Within this framework, the \ac{PSD} and cross-spectral density of the frequency-domain signal $\widetilde{h}_{I,J}$ are determined by
\be
\label{eq:hIhJ}
\langle\widetilde{h}_{I}(f,\tau)\widetilde{h}_{J}^{*}(f',\tau)\rangle
=\frac{1}{2}\delta(f-f')\Gamma_{IJ}(f,\tau)S_{\rm h}(|f|),\\
\ee
where the \ac{ORF} involves both the channel response $F_{I}^{P}$ and the separation vector $\Delta \vec{x}=\vec{x}_{I}-\vec{x}_{J}$ between channels\footnote{For space-borne detectors, a satellite can be selected as the reference location of the channel.}~\cite{Liang:2022ufy}:
\bea
\label{eq:Gamma_IJ}
\nn
\Gamma_{IJ}(f,\tau)=&&
\frac{1}{8\pi}\sum_{P}
\int_{S^{2}}{\rm d}\hat{\Omega}_{\hat{k}}\,
F^{P}_{I}(f,\hat{k},\tau)F^{P*}_{J}(f,\hat{k},\tau)\\
&&\times
e^{-{\rm i}2\pi f\hat{k}\cdot \Delta \vec{x}(\tau)/c}.
\eea

In the \ac{SGWB} detection, it is imperative to account for not only the signal itself but also the detector noise. 
For space-borne detectors, the inherent motion introduces uncancelable phase noise. It is necessary to use the \ac{TDI} techniques to suppress the phase noise by several orders of magnitude, ensuring that the phase noise is lower than the target \ac{GW} level~\cite{Tinto:1999yr,Tinto:2020fcc}. 
Among the various \ac{TDI} combinations, the unequal-arm Michelson (X, Y, Z) involves the measurements of neighboring links for each satellite. 
Building upon the (X, Y, Z) setup, a more advanced setup referred to as the noise-orthogonal unequal-arm Michelson (A, E, T) can be constructed, according to Ref.~\cite{Vallisneri:2007xa}, as follows\footnote{In this paper, we assume a completely equilateral configuration; otherwise, the AET channels will not be noise-orthogonal~\cite{Hartwig:2023pft}.}:
\bea
\label{eq:channel_AET}
\nn
{\rm A}&=&\frac{1}{\sqrt{2}}({\rm Z}-{\rm X}),\\
\nn
{\rm E}&=&\frac{1}{\sqrt{6}}({\rm X}-2{\rm Y}+{\rm Z}),\\
{\rm T}&=&\frac{1}{\sqrt{3}}({\rm X}+{\rm Y}+{\rm Z}).
\eea
Assuming that the noise is stationary, the noise \ac{PSD} for the (A, E, T) setup is given by
\bea
\label{eq:Pn_AET}
\nn
P_{\rm n_{\rm A/E}}(f)
&=&
\frac{2\sin^{2}u}{L^{2}}
\bigg[\big(\cos u+2\big)S_{\rm p}(f)\\
\nn
&&+2\big(\cos(2 u)+2\cos u+3\big)
\frac{S_{\rm a}(f)}{(2\pi f)^{4}}\bigg],\\
P_{\rm n_{T}}(f)
&=&
\nn
\frac{8\sin^{2}u
\sin^{2}\frac{u}{2}}{L^{2}}
\bigg[S_{\rm p}(f)+4\sin^{2}\frac{u}{2}
\frac{S_{\rm a}(f)}{(2\pi f)^{4}}\bigg],\\
\eea
where $u=(2\pi fL)/c$ is determined by the detector arm length $L$, $S_{\rm p}$ and $S_{\rm a}$ denote the \acp{PSD} of optical-metrology system noise and acceleration noise, respectively. 
For more comprehensive details about the parameters of TianQin and LISA, readers are recommended to refer to Refs.~\cite{TianQin:2020hid,Babak:2021mhe}.
Unless otherwise stated, our study primarily focuses on the case using the (A, E, T) setup.

\section{Methodology}\label{sec:method}
Statistical inference is a powerful tool for identifying the presence of an \ac{SGWB} signal within observation data.  
This process is commonly pursued through both classical (frequentist) and Bayesian inference methodologies. 
In this section, we constrain our focus on parameter estimation and model selection in the context of \ac{SGWB} detection, opting for a Bayesian inference approach.

\subsection{Bayesian inference}
Bayesian inference offers a robust quantitative framework for gauging the uncertainties associated with unknown parameters. 
This estimation process hinges on the posterior probability distribution, which is formulated through the application of Bayes' theorem. 
Bayes' theorem updates the prior distribution $p(\boldsymbol{\theta})$ of the parameters $\boldsymbol{\theta}$ with the likelihood $p(d|\boldsymbol{\theta})$, thereby yielding the posterior distribution of the observations given the parameters:
\be
\label{eq:pot_dib}
p(\boldsymbol{\theta}|d)=
\frac{p(d|\boldsymbol{\theta})p(\boldsymbol{\theta})}{p(d)},
\ee
where the Bayesian evidence serves as a normalization factor:
\be
p(d)=\int {\rm d \boldsymbol{\theta}}\,
p(d|\boldsymbol{\theta})p(\boldsymbol{\theta}).
\ee
The posterior distribution encodes the information from the data with our prior knowledge, providing a comprehensive estimation of the parameters.

Bayesian inference is not merely for parameter estimation; it also serves as a discerning tool for selecting a credible model or hypothesis from a collection of alternatives, each characterized by a set of parameters. 
In the subsequent discussion, we will denote the hypothesis as $\mathcal{H}_{a}$, where the subscript index $a$ encompasses the range of competing hypotheses and the associated set of parameters as $\boldsymbol{\theta}_{a}$. 
Regarding~\eq{eq:pot_dib}, the joint posterior distribution for the parameters $\boldsymbol{\theta}_{a}$, conditioned on the $\mathcal{H}_{a}$, is expressed as
\be
p(\boldsymbol{\theta}_{a}|d,\mathcal{H}_{a})=
\frac{p(d|\boldsymbol{\theta}_{a},\mathcal{H}_{a})p(\boldsymbol{\theta}_{a}|\mathcal{H}_{a})}{p(d|\mathcal{H}_{a})}.
\ee
In terms of the evidence $p(d|\mathcal{H}_{a})$, the posterior probability for $\mathcal{H}_{a}$ is provided by Bayes' theorem as follows:
\be
p(\mathcal{H}_{a}|d)=\frac{p(d|\mathcal{H}_{a})p(\mathcal{H}_{a})}{P(d)},
\ee
where $P(d)$ denotes the evidence across all the models considered. 
To compare the relative support $\mathcal{H}_{a}$ and $\mathcal{H}_{b}$, the posterior odds ratio is computed:
\be
\mathcal{O}_{ab}(d)=\frac{p(\mathcal{H}_{a}|d)}{p(\mathcal{H}_{b}|d)}=\frac{p(\mathcal{H}_{a})}{p(\mathcal{H}_{b})}\frac{p(d|\mathcal{H}_{a})}{p(d|\mathcal{H}_{b})}.
\ee
In the absence of a priori preferences favoring one hypothesis over the other, the prior probabilities $p(\mathcal{H}_{a})$ and $p(\mathcal{H}_{b})$ are assumed equal, reducing the odds ratio to the {\it Bayes factor}:
\be
\mathcal{B}_{ab}=\frac{p(d|\mathcal{H}_{a})}{p(d|\mathcal{H}_{b})}.
\ee
The Bayes factor thus emerges as a critical metric for model selection. 
A positive log-Bayes factor, $\ln (\mathcal{B}_{ab}$), signifies that the data favors the hypothesis $\mathcal{H}_{a}$. 
It is conventional to interpret $\ln \mathcal{B}_{ab}>1$ as an indication of positive evidence, a stronger level of support for the model $\mathcal{H}_{a}$ is acknowledged when $\ln \mathcal{B}_{ab}>3$, and the support is considered very strong when $\ln\mathcal{B}_{ab}>5$~\cite{Robert:BF}.

Nested sampling~\cite{dynested} is an advanced technique used to provide samples of the posterior distribution, given a specified likelihood function and prior distribution~\cite{Skilling:2006gxv,Higson:2018cqj,Speagle:2019ivv,Ashton:2022grj}. 
The algorithm starts by randomly selecting a fixed number of live points from the prior distribution. 
In each iteration $i$, the live point with the lowest likelihood is replaced with a new point drawn from the prior region where the likelihood $p(d|\boldsymbol{\theta})$ exceeds $p_{i}$. 
The iterative process continues until a predefined termination criterion is met, yielding a collection of samples (referred to as dead points) along with any remaining live points. 
These points are then employed for both evidence calculation and parameter estimation, where the evidence can be utilized for model selection.

\subsection{Likelihood for cross-correlation detection}\label{subsec:llh}
The analysis presented above has indicated that the likelihood function is of critical importance for parameter estimation and model selection. 
Next, we turn our attention to derive the likelihood for the \ac{SGWB} detection.

Given the constraints of the null-channel method for single detectors, which are significantly limited by uncertainties in noise knowledge~\cite{Muratore:2023gxh}, we propose a cross-correlation approach for \ac{SGWB} detection using the TL network. 
However, the variability of the \ac{ORF} for the TL network necessitates segmenting the data into manageable intervals. 
Each segment must be longer than the light-travel time between the detectors to ensure effective correlation, yet short enough to maintain the stability of the \ac{ORF} throughout. 
For each segment, a Fourier transform over the designated time interval can be given by
\be
\widetilde{s}^{i}_{I}(f)=\int_{t_{i}-T/2}^{t_{i}+T/2}{\rm d}t \,
s_{I}(t)e^{-{\rm i}2\pi ft},
\ee
where $i$ labels each time interval. 
By utilizing the Fourier transforms $\widetilde{s}^{i}_{I}$, one can construct the following estimator for the $S_{\rm h}$ measured by detectors~\cite{LIGOScientific:2014sej}:
\be
\label{eq:S_IJ}
\hat{C}^{i}_{IJ}(f)
=\frac{2}{T}
\frac{\widetilde{s}^{i}_{I}(f)\widetilde{s}^{i*}_{J}(f)}{\Gamma^{i}_{IJ}(f)}.
\ee
Assuming that the detector noise is Gaussian, stationary, the variance of the estimators is given by\footnote{While this formula applies to the \ac{SGWB} of any intensity, it is crucial to emphasize that the \ac{ORF} for cross-correlation between two detectors should be significantly less than that for auto-correlation within single detectors. The TL network, as noted in Ref.~\cite{Liang:2024ulf}, adheres to this essential criterion.}
\be
\label{eq:sigma_IJ}
\big(\sigma_{IJ}^{i}(f)\big)^{2}\approx
\frac{1}{2\,T\delta f}\frac{P_{{\rm s}_{I}}(f)P_{{\rm s}_{J}}(f)}{|\Gamma^{i}_{IJ}(f)|^{2}},
\ee
where $\delta f$ is the width of each frequency bin, the total \ac{PSD}
\be
\label{eq:P_tot}
P_{{\rm s}_{I}}(f)=P_{{\rm n}_{I}}(f)+\Gamma_{II}(f)S_{\rm h}(f).
\ee
It is crucial to recognize that in the weak-signal limit, where the \ac{SGWB} signal is significantly weaker than the detector noise, the total \ac{PSD} can be attributed entirely to the detector noise. 
However, for space-borne detectors such as TianQin and LISA, the anticipated \ac{SGWB} signal can surpass the detector noise.  Consequently, the \ac{PSD} as described in~\eq{eq:P_tot} should be amended to include the contribution of the \ac{SGWB} signal.

For the TL network, cross-correlation can be effectively carried out using the four-channel pairs $\{\rm AA', AE', EA', EE'\}$~\cite{Seto:2020mfd,Liang:2022ufy}, where the prime symbol is used to differentiate between TianQin and LISA. 
Consequently, when the data is segmented into $N$ parts, there are $4N$ $\hat{C}_{IJ}$ to handle. 
To achieve an optimal expected \ac{SNR} for \ac{SGWB} detection, the numerous $\hat{C}_{IJ}(f)$ can be combined into a single estimator through a weighted sum~\cite{Allen:1997ad,Callister:2016ewt,Callister:2017ocg,Hu:2023nfv}:
\be
\label{eq:C_tot}
\hat{C}_{\rm tot}(f)
=\frac{\sum_{i}\sum_{IJ}\hat{C}^{i}_{IJ}(f)\big(\sigma_{IJ}^{i}(f)\big)^{-2}}{\sum_{i}\sum_{IJ}\big(\sigma_{IJ}^{i}(f)\big)^{-2}}.
\ee
This construction process is analogous to data folding~\cite{Ain:2015lea,Ain:2018zvo}, which helps in efficiently handling the data.

According to the construction, the weighted estimator is also applicable to the $S_{\rm h}$:
\bea
\nn
\langle\hat{C}_{\rm tot}(f)\rangle 
&=&\frac{\langle\hat{C}^{i}_{IJ}(f)\rangle\sum_{i}\sum_{IJ}\big(\sigma_{IJ}^{i}(f)\big)^{-2}}{\sum_{i}\sum_{IJ}\big(\sigma_{IJ}^{i}(f)\big)^{-2}}\\
&=&S_{\rm h}(f),
\eea
and the variances of the optimal estimator are added in the same way as electrical resistors in parallel to maximize \ac{SNR}~\cite{Allen:1997ad}:
\bea
\nn
\label{eq:sigma_tot}
\sigma^{2}_{\rm tot}(f)&=&
\frac{1}{\sum_{i}\sum_{IJ}\big(\sigma_{IJ}^{i}(f)\big)^{-2}}\\
\nn
&\approx&
\frac{1}{2\,T\delta f \sum_{i}\sum_{IJ}|\Gamma^{i}_{IJ}(f)|^{2}/\big(P_{{\rm s}_{I}}(f)P_{{\rm s}_{J}}(f)\big)}
\\
&=&
\frac{1}{2\,T\delta f}\frac{P_{{\rm s}_{I}}(f)P_{{\rm s}_{J}}(f)}{N\bar{\Gamma}^{2}_{\rm tot}(f)},
\eea
where it is assumed that the channels of the same detector share the same noise \ac{PSD}, allowing for a more straightforward calculation of the total uncertainty in the cross-correlation. 
The time-averaged \ac{ORF} is given by
\be
\bar{\Gamma}_{\rm tot}(f)
=\sqrt{\frac{\sum_{i}\sum_{IJ}|\Gamma^{i}_{IJ}(f)|^{2}}{N}}.
\ee
In the context of measuring the specific $\hat{C}_{\rm tot}(f)$ within a single frequency bin, the log-likelihood is expressed as
\be
\label{eq:llh_f}
\ln \left[p(\hat{C}_{\rm tot}(f)|\boldsymbol{\theta})\right]
\propto -\frac{|\hat{C}_{\rm tot}(f)-S_{\rm h}(f,\boldsymbol{\theta})|^{2}}{2\,\sigma^{2}_{\rm tot}(f)},
\ee
where $\boldsymbol{\theta}$ denotes the model parameter of the \ac{SGWB}, and the variance is provided by~\eq{eq:sigma_tot}. 
For the entire frequency range, the full log-likelihood is the sum of individual log-likelihoods across each frequency bin:
\be
\label{eq:llh}
\ln \left[\mathcal{L}(\boldsymbol{\theta})\right]
=\mathcal{N} \sum_{f} \ln \left[p(\hat{C}_{\rm tot}(f)|\boldsymbol{\theta})\right],
\ee
where $\mathcal{N}$ is a normalization factor. 

\section{Analysis setup}\label{sec:setup}
The full log-likelihood presented in~\eq{eq:llh} involves several critical aspects: the cross-correlation estimator $\hat{C}_{\rm tot}$, the models of the \ac{SGWB} intensity ($S_{\rm h}$ or $\Omega_{\rm gw}$), the priors of model parameters, and the corresponding statistical error $\sigma_{\rm tot}$. 
Next, we will undertake a detailed examination of each of these elements, aiming to construct a clear and coherent narrative that elaborates the data processing pipeline for cross-correlation detection. 

\subsection{Segment duration}
\begin{figure*}[t]
     \centering
     \includegraphics[width=.80\linewidth]{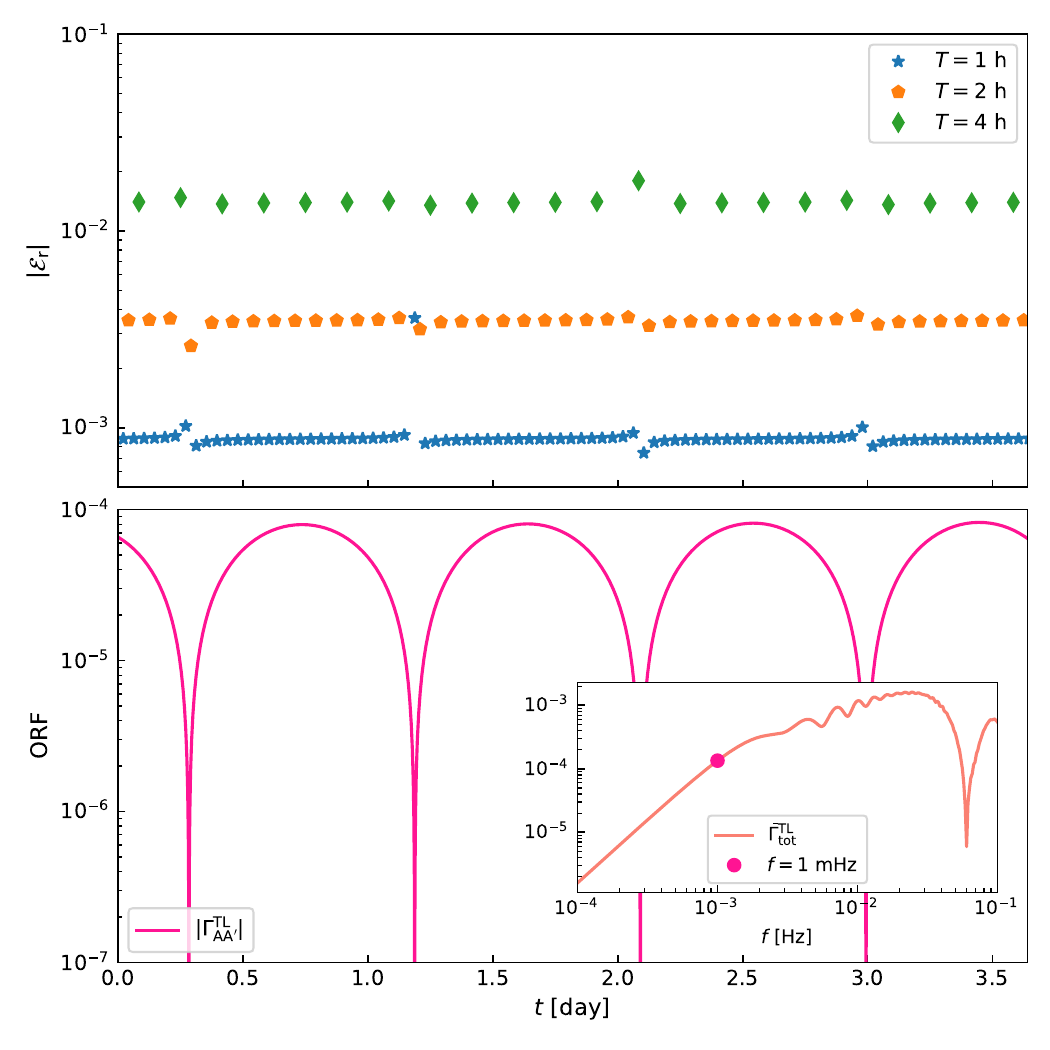}
     \caption{Relative error of the \ac{ORF} at 1 mHz. In the top panel, the points are color-coded to correspond to different segment lengths: blue, orange, and green dots denote 1 hour, 2 hours, and 4 hours, respectively. In the bottom panel, the results of the \ac{ORF} are also included for reference, with the total \ac{ORF} for $\{\rm AA', AE', EA', EE'\}$ presented in the inset.}
     \label{fig:Error_orf_t}
\end{figure*}

In the construction of the cross-correlation estimator, calculating the \acp{ORF} for each segment is essential. 
If the \ac{ORF} does not change significantly within a segment, one can use the midpoint \ac{ORF} value as a representative for the entire segment.
Nonetheless, the effective \ac{ORF} for this segment should be derived by dividing the \ac{ORF} integral over that time period by the data duration $T$:
\be
\Gamma_{IJ}^{\rm eff}(f)=\frac{1}{T}
\int_{-T/2}^{T/2}{\rm d}t\,\Gamma_{IJ}(t,f).
\ee
This approximation can impact the accuracy and reliability of the subsequent Bayesian inference. 
Shorter segments inherently reduce the deviation, but practical considerations must be taken into account: to ensure the integrity of frequency-domain data and prevent the loss of information within the sensitive band of the detectors, it is crucial to select segments of sufficient duration. 
Both TianQin and LISA are sensitive within the mHz frequency band, necessitating the choice of at least an hourly segment duration.

Given the significantly shorter orbital period of TianQin compared to \ac{LISA}, it can be approximated that \ac{LISA} remains stationary while TianQin completes its 3.64-day orbital period. 
Moreover, given that the A/E channels can be effectively considered as two L-shaped interferometers with an offset angle of $45\degree$~\cite{Seto:2020mfd}, we can expect that the period of the ORF for $\{\rm AA', AE', EA', EE'\}$ will be one-fourth (as TianQin rotates $90\degree$) of that of TianQin, and we can choose the \ac{ORF} for the $\rm AA'$ pair as a representative case.
In~\fig{fig:Error_orf_t}, we present the absolute value of the relative error $\mathcal{E}_{\rm r}$ of the \ac{ORF} $\Gamma^{\rm TL}_{\rm AA'}$ at 1 mHz~\footnote{In our previous paper~\cite{Liang:2022ufy}, it can be found that the most sensitive frequency range for the TL network is approximately between 0.5 and 5 mHz. Consequently, we utilize 1 mHz as a reference frequency.}, over a specific period of 3.64 days. 
This error is derived by comparing the effective value to a representative value at the midpoint for \ac{ORF} within each segment. 
For context and clarity in our analysis, we also depict the \ac{ORF} $\Gamma^{\rm TL}_{\rm AA'}$ itself, along with the time-averaged \ac{ORF} $\bar{\Gamma}^{\rm TL}_{\rm tot}$ of the TL network\footnote{For detailed calculations of \ac{ORF}, readers can refer to our previous series of work, e.g.,~Refs.~\cite{Liang:2021bde,Liang:2022ufy,Liang:2023fdf}.}. 
Our analysis reveals that as the segment duration expands from 1 hour to 4 hours, $|\mathcal{E}_{\rm r}|$ rises from 0.1\% to 1\%. 
However, the period of ORF is a quarter of the period of TianQin, which is approximately 22 hours. In such a situation, if the segment duration is set to 1 hour, the contribution of the ORF (close to 0) for this small piece to the effective ORF is insignificant. Even if this portion of the data is lost, it will merely have a rather minor influence on our present analysis.
Based on the above analysis, we will adopt a segment duration of 1 hour. 

\subsection{Components of stochastic background}\label{sec:components}
In the mHz band, data collected by space-borne detectors has the potential to include the Galactic foreground~\cite{Ruiter:2007xx,Cornish:2017vip,Huang:2020rjf,Liang:2021bde}. 
As the observation time extends, a growing number of individual \acp{DWD} can exceed the predetermined detection threshold. 
The identification and subsequent removal of these \acp{DWD} will reduce the intensity of the Galactic foreground. 
In our previous work, we established that TianQin and the \ac{LISA} will be capable of simultaneous operation for up to 4 years~\cite{Liang:2021bde}. 
Compared with the observation of a single detector, the joint observation of the two detectors enables more resolvable \acp{DWD}, and thus the foregrounds for TianQin and the LISA can be further depressed, respectively. Based on the two foregrounds, a joint foreground of the TL network can be derived through Eq. (41) of Ref.~\cite{Liang:2021bde}.
For the joint foreground, corresponding polynomial coefficients $a_{i}$ required for~\eq{eq:Omega_fg} are specified in~\tab{tab:DWD}. 
\begin{table}
	\begin{center}
		\caption{Coefficients for the polynomial fit for the Galactic foreground. As per the original plans for the TianQin and LISA missions, they can be scheduled for a joint operation period of up to 4 years~\cite{Liang:2021bde}. During this operation time $T$, the foreground is anticipated to be depressed as the accumulated observation time grows. Note that, due to TianQin's working mode, it only offers a dataset with a duration of $T/2$ during the operation period. For example, if the operation time is 0.5 years, the duration of the dataset will be 3 months.}
		\centering
		\setlength{\tabcolsep}{1.4mm}
		\renewcommand\arraystretch{1.5}
		\label{tab:DWD}
		\begin{tabular}{c c c c c c c c}
			\hline
			\hline
			$T$   &$a_{0}$ & $a_{1}$&$a_{2}$ &$a_{3}$& $a_{4}$&$a_{5}$ &$a_{6}$\\
			\hline
			0.5 yr  & -18.71  & -1.41   & -0.550  &  -2.42  &  2.86    &  10.0   & -19.9\\
			1 yr    & -18.74  & -1.21   & -0.613  &  -3.07  &  1.89    &  3.15   & -17.4\\
			2 yr    & -18.75  & -1.47   & -1.44   &  -0.568 & 2.07    &  -1.21  & -21.1 \\
			4 yr    & -18.79  & -1.49   & -0.895  &  -3.58  & -8.28   &  1.32   & -1.44\\
			\hline
			\hline
		\end{tabular}
	\end{center}
\end{table}
\begin{figure}[t]
	\centering
	\includegraphics[width=.95\linewidth]{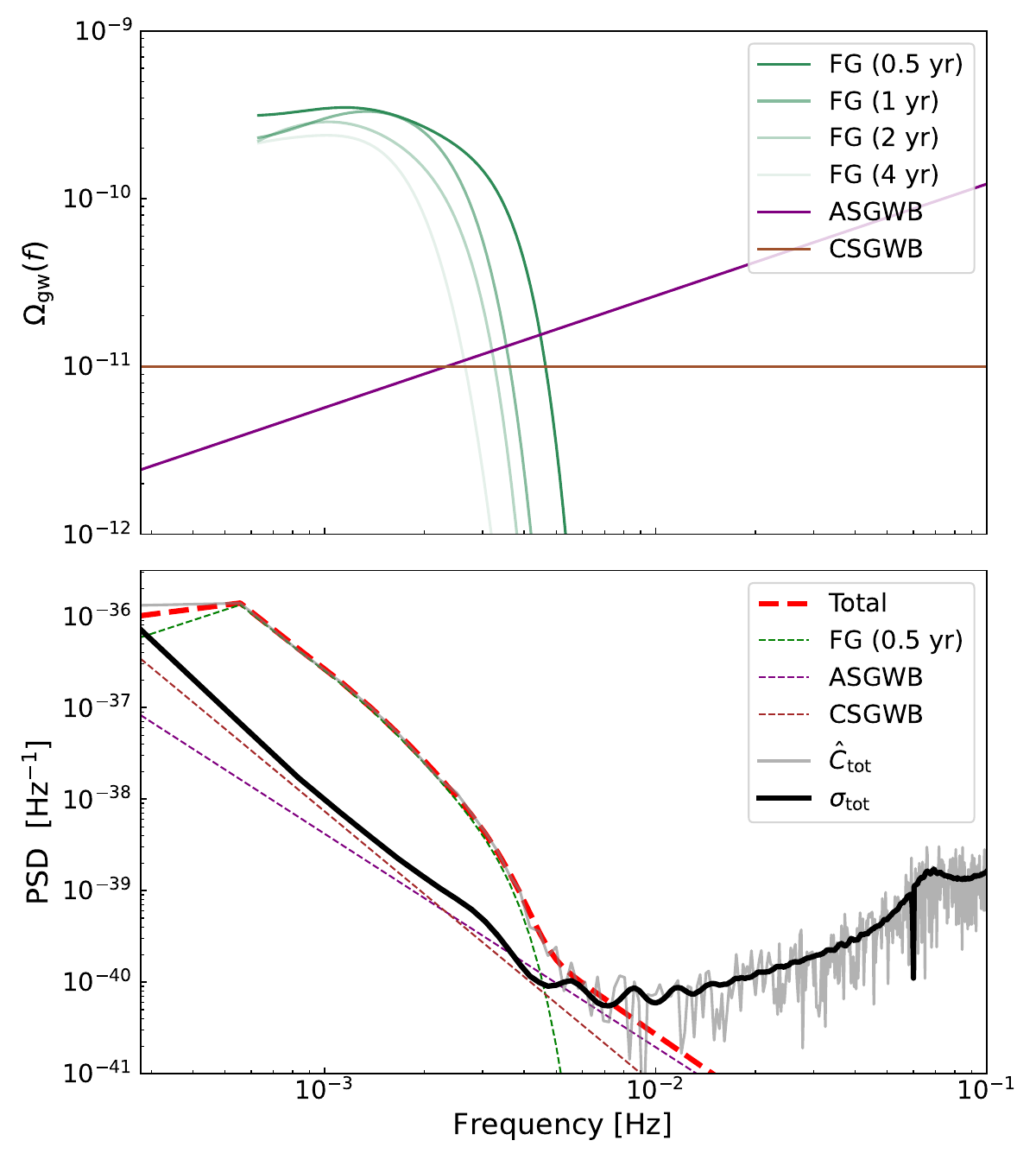}
	\caption{Top panel: energy spectrum density of the Galactic foreground corresponding to different operation times (green), the astrophysical background (purple), and the cosmological background (brown). Bottom panel: estimator $\hat{C}_{\rm tot}$ (gray solid line) and standard deviation $\sigma_{\rm tot}$ (black solid line) for the 3-month simulated data of the TL network, where we approximate the total background (red dashed line) as a collection of 0.5-year Galactic foreground (green dashed line), astrophysical background (purple dashed line), and cosmological background (brown dashed line). Note that in the frequency band below 0.6 mHz, the polynomial-fitting results will plunge to 0 and no longer match the intensity of the actual Galactic foreground. Nevertheless, in this frequency band, the foreground is far lower than the detector noise. Thus, we have excluded it from the top panel. Regarding the bottom panel, to obtain the discrete data points, the Galactic foreground needs to extend to 1/3600 Hz, and two points seem to form a straight line at low frequencies.}
	\label{fig:SGWB_tot}
\end{figure}

In~\fig{fig:SGWB_tot} (top panel), we show the energy spectrum density $\Omega_{\rm gw}$ of the foreground using green lines. 
Although the foreground dominates the frequency band around mHz, its intensity sharply declines at a specific frequency. 
This drop frequency shifts from approximately 5 to 3 millihertz as the operation time increases, thereby revealing other backgrounds at higher frequencies that were previously masked. 
These exposed backgrounds can originate from the extragalactic binaries of astrophysical origin and the cosmic defects of cosmological origin~\cite{Liang:2021bde}. 
For the astrophysical \ac{SGWB}, the value of $\Omega_{\rm ast}$ as defined in~\eq{eq:Omega_ast} is anticipated to fall within the range $[5.28\times10^{-12},1.32\times10^{-10}]$ when using a reference frequency $f_{\rm ref}=10\,\,{\rm mHz}$~\cite{Liang:2021bde}. 
Regarding the cosmological \ac{SGWB}, the $\Omega_{\rm ast}$ as specified in~\eq{eq:Omega_cos} is constrained to be below $10^{-10}$. 
As illustrated in the top panel of~\fig{fig:SGWB_tot}, the purple line represents the astrophysical background with $\Omega_{\rm ast}=2.64\times10^{-11}$, while the brown line denotes the cosmological background with $\Omega_{\rm cos}=1\times10^{-11}$. 
In the bottom panel, the \acp{PSD} of these backgrounds are illustrated, along with the 0.5-year Galactic foreground. 
Considering the total background, which includes these three components, we simulate 3-month data for the channel group $\{\rm A, E, A', E'\}$ of TianQin and LISA. 
By cross-correlating the data from four channel pairs $\{\rm AA', AE', EA', EE'\}$, we construct the estimator $\hat{C}_{\rm tot}$ as defined in~\eq{eq:C_tot}. 
This estimator, along with the statistical error $\sigma_{\rm tot}$, is also shown in the bottom panel. 
We find that the contribution of \acp{SGWB} to the estimator only surpasses that of the Galactic foreground and statistical error within the frequency range of 5 to 6 mHz. 
To broaden this advantageous range, it is essential to depress the Galactic foreground and statistical error by extending the operation time. 
It should be noted that for component separation, we only concentrate on the relative amplitude of the \ac{SGWB} to the Galactic foreground. Furthermore, the peculiar long-term modulation and non-Gaussianities of the Galactic foreground can also enable component separation~\cite{Pozzoli:2024wfe,Buscicchio:2024wwm,Karnesis:2024pxh,Rosati:2024lcs}.

\subsection{Priors}
In selecting the priors for the parameter, we opt for a weakly constrained range to ensure a robust analysis. 
For the Galactic foreground, the impact of higher-order polynomial coefficients $a_{i}$ on its intensity becomes more pronounced. 
Taking into account the variations in $a_{i}$ for each order across different operation times, we establish the prior ranges for these coefficients as follows:
\bea
a_{i}\in\left\{
\begin{array}{lr}
	\left[a_{i,{\rm true}}-1,a_{i,{\rm true}}+1\right],\,\, i=0,1,2  \\
	\\
	\left[a_{i,{\rm true}}-5,a_{i,{\rm true}}+5\right],\,\, i=3,4
	\\
	\\
	\left[a_{i,{\rm true}}-10,a_{i,{\rm true}}+10\right],\,\, i=5,6
\end{array},
\right.
\eea
where $a_{i,{\rm true}}$ represents the true value. 

Regarding the other \ac{SGWB} parameters, we impose uniform priors on $\log(\Omega_{\rm ast,cos})$ and $\alpha_{\rm ast}$ within a range extending two units above or below the true value, i.e., the interval $[\log(\Omega_{\rm ast,cos})-2,\log(\Omega_{\rm ast,cos})+2]$ and $[\alpha_{\rm ast}-2,\alpha_{\rm ast}+2]$.

\subsection{Total PSD estimation}
\begin{figure}[h]
	\centering
	\includegraphics[width=.95\linewidth]{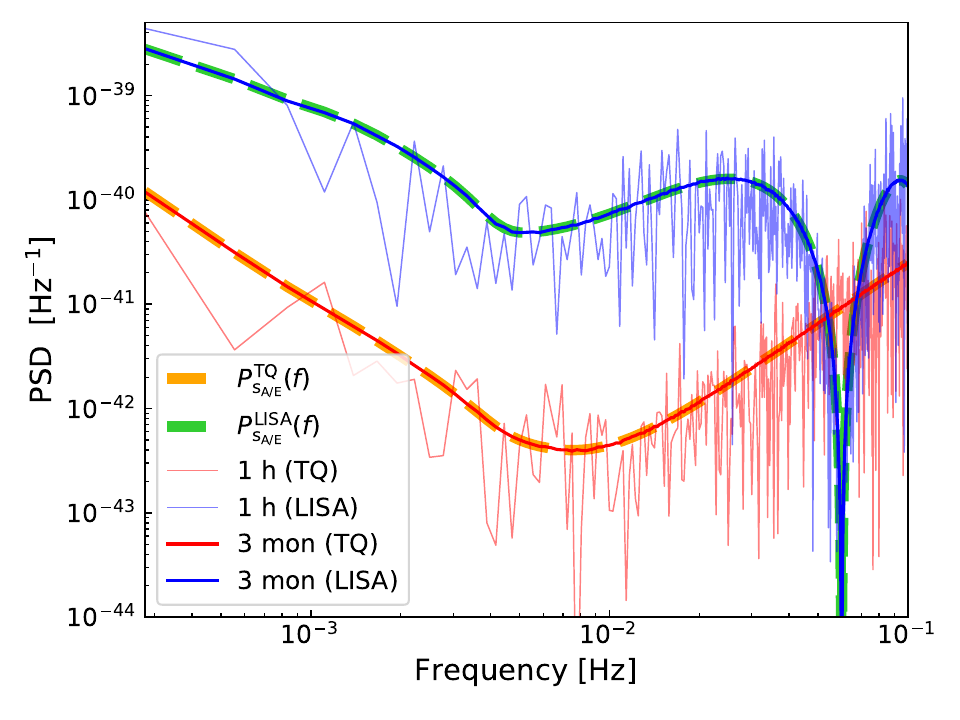}
	\caption{Estimation of the total \ac{PSD}. The thick dotted orange and green lines correspond to the true \acp{PSD} of TianQin and LISA, respectively. The solid red and blue lines depict the estimated \acp{PSD} for TianQin and LISA, derived from 1-hour and 3-month auto-correlation data.}
	\label{fig:P_tot}
\end{figure}

\begin{figure*}[t]
	\centering
	\includegraphics[width=.95\linewidth]{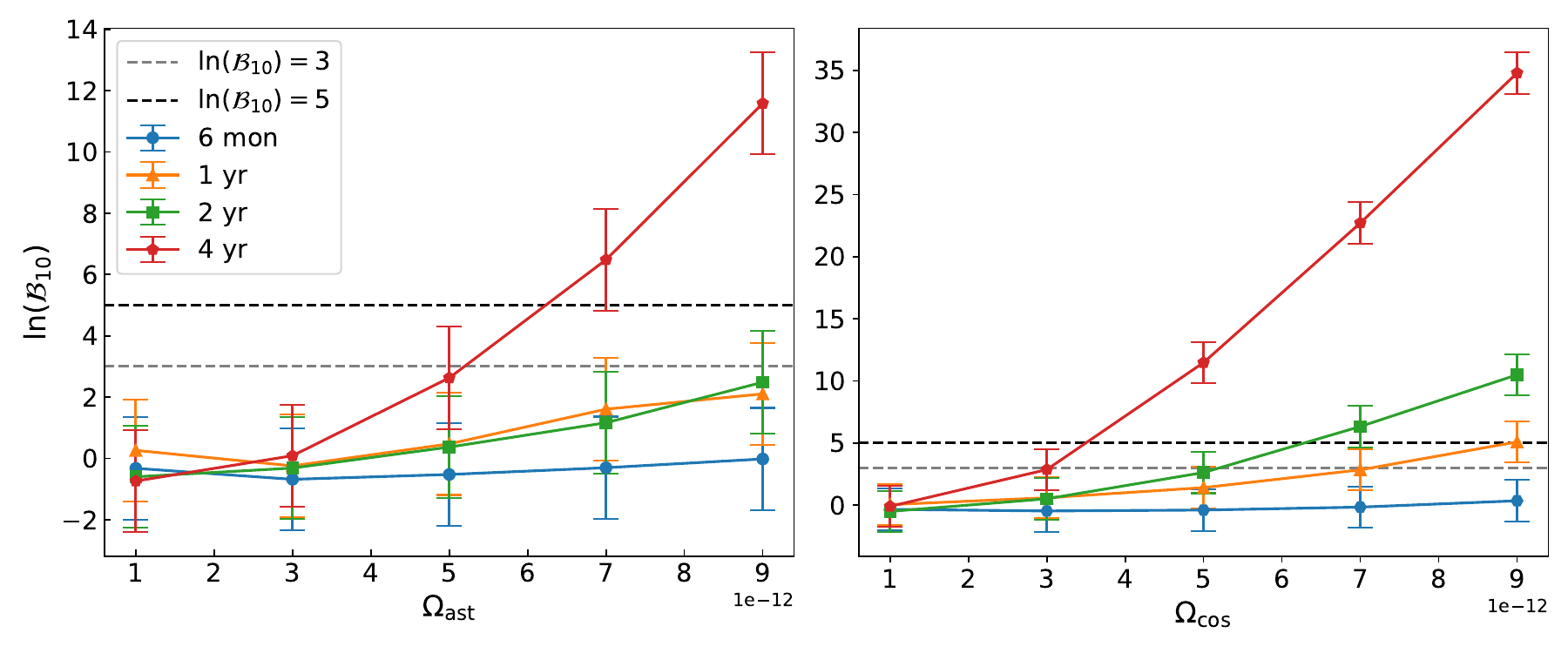}
	\caption{Log-Bayes factor, between $\mathcal{H}_{0}$, $\mathcal{H}_{1}$, as a function of astrophysical energy density $\Omega_{\rm ast}$ (with a variable index $\alpha_{\rm ast}$, left panel) and cosmological energy density $\Omega_{\rm cos}$ (with a fixed index, right panel). The blue, orange, green, and red lines correspond to the TL network's performance over operation time of 0.5, 1, 2, and 4 years, respectively. The error bars indicate the quantiles [16\%, 84\%], with the points representing the median. For reference, the dashed gray lines indicate log-Bayes factors of 3 and 5, serving as thresholds for the detection of an additional \ac{SGWB} in the presence of Galactic foreground.} 
	\label{fig:B_ast_cos}
\end{figure*}

\begin{figure}[t]
	\centering
	\includegraphics[width=.95\linewidth]{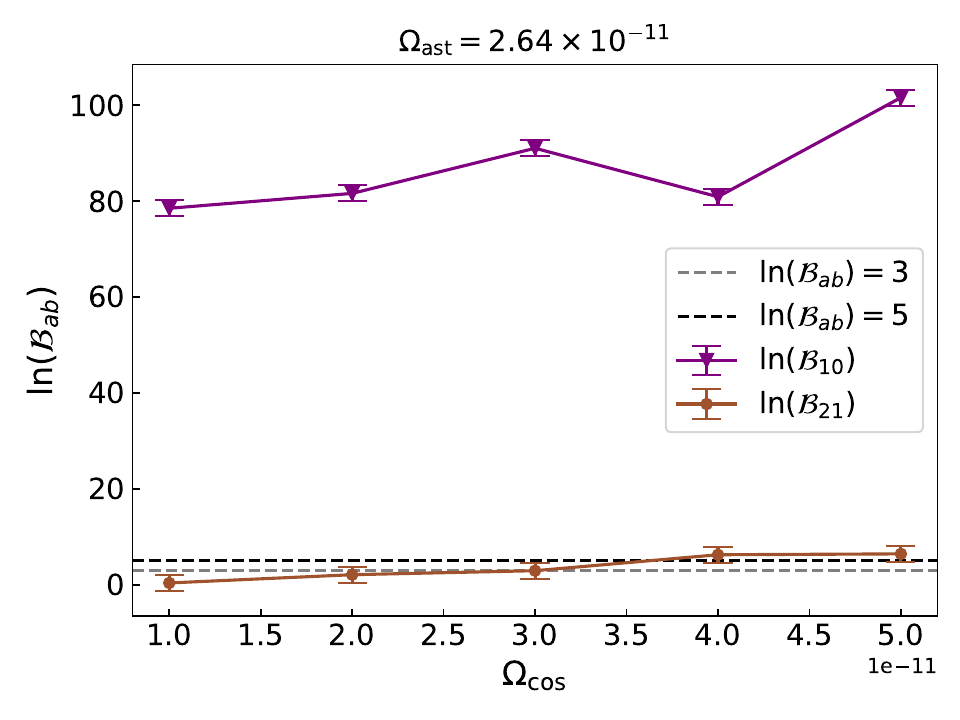}
	\caption{Log-Bayes factors $\ln (\mathcal{B}_{10})$ (purple) and $\ln (\mathcal{B}_{21})$ (brown) for hypotheses $\mathcal{H}_{0}$, $\mathcal{H}_{1}$, and $\mathcal{H}_{2}$, assuming an operation time of 4 years. The error bars indicate the quantiles [16\%, 84\%], with the points representing the median. The dashed gray lines represent the log-Bayes factor thresholds of 3 and 5, respectively.}
	\label{fig:B_tot}
\end{figure}

To derive the statistical error $\sigma_{\rm tot}$ for likelihood, as shown in~\eq{eq:sigma_tot}, one needs to model the total \ac{PSD} $P_{{\rm s}}$ of the data. 
In this work, we estimate the total \ac{PSD} by auto-correlating the data, an approach that aligns with the generation of cross-correlation data as outlined in~Sec.~\ref{subsec:llh}. 
Starting from~\eq{eq:S_IJ}, the auto-correlation estimator for a given time interval can be derived by setting $I=J$ and $\Gamma_{IJ}=1$. 
This estimator is inherently subject to statistical uncertainties. 
Nevertheless, by employing~\eq{eq:C_tot}, which entails averaging multiple auto-correlation data across each frequency bin, these uncertainties can be substantially reduced by a factor of $1/N$, as evidenced in~\eq{eq:sigma_tot}. 
Assuming that the \ac{PSD} is both smooth~\cite{Welch:1967} and stationary\footnote{A more practical scenario is that a noise PSD is unable to remain stationary~\cite{LISAPathfinder:2024ucp}.}, a sufficiently large number of data segments $N$ ensures that the above procedure can reliably infer the shape of the total \ac{PSD}.

For the A channel of TianQin and LISA, we simulate the corresponding 3-month datasets, encompassing detector noise, Galactic foreground corresponding to an operation time of 0.5 years, and the other components of the \ac{SGWB} with the energy spectrum density $\Omega_{\rm gw}$ as depicted in~\fig{fig:SGWB_tot}.  
This dataset is subsequently divided into 2190 segments, each lasting 1 hour. 
As illustrated in~\fig{fig:P_tot}, the \ac{PSD} derived directly from a single 1-hour auto-correlation data segment, without the benefit of averaging, exhibits a pronounced deviation from the true \ac{PSD}. 
In sharp contrast, averaging over all 2190 data segments dramatically reduces the statistical uncertainty, yielding an estimated \ac{PSD} that is closely aligned with the true \ac{PSD}. 
We conclude that by dividing and averaging the 3-month auto-correlation data into 1-hour segments, it is feasible to derive total \acp{PSD}, allowing us to effectively estimate the statistical error associated with the cross-correlation estimator.

\begin{figure*}[t]
	\centering
	\includegraphics[width=.80\linewidth]{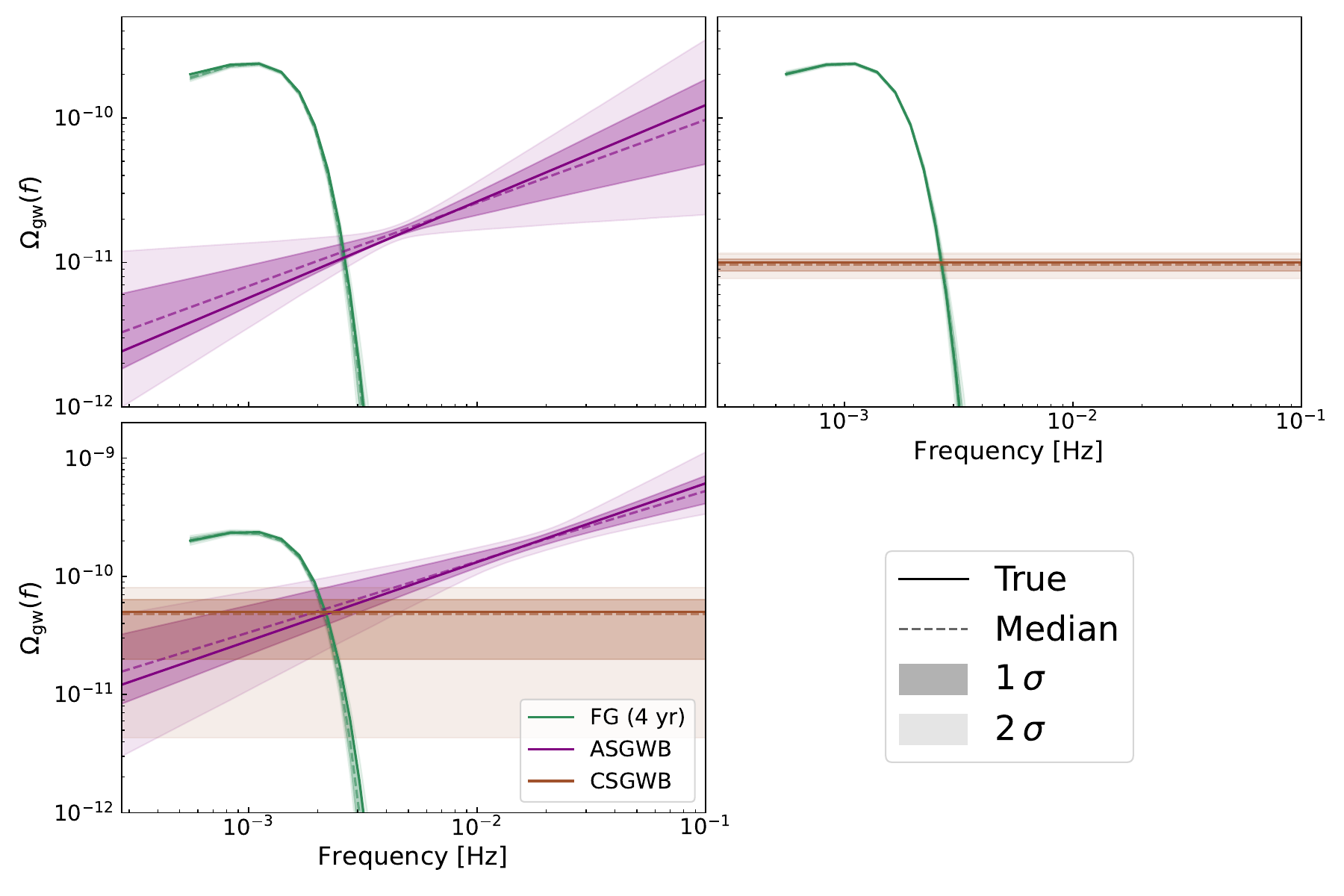}
	\caption{Posteriors of energy spectrum densities $\Omega_{\rm gw}$ of Galactic foreground and \acp{SGWB}, with the 4-year operation of the TL network. The green, purple, and brown lines represent the Galactic foreground, the astrophysical background, and the cosmological background, respectively. The posteriors of the Galactic foreground are related to seven polynomial coefficients $a_{i}$, those of the astrophysical background to $\Omega_{\rm ast}$ and $\alpha_{\rm ast}$, and those of the cosmological background to $\Omega_{\rm cos}$. The true and estimated median values are distinguished by solid and dashed lines. The lightly shaded areas indicate the 1-$\sigma$ and 2-$\sigma$ credible regions. Both panels include a 4-year Galactic foreground: the top panel presents the parameter estimation on the astrophysical and cosmological backgrounds with $\Omega_{\rm ast}=2.64\times10^{-11}$ and $\Omega_{\rm cos}=1\times10^{-11}$, respectively; the bottom panel illustrates the parameter estimation on both the astrophysical and cosmological backgrounds, where $\Omega_{\rm ast}=1.32\times10^{-10}$ and $\Omega_{\rm cos}=5\times10^{-11}$.}
	\label{fig:PE_tot}
\end{figure*}
\section{Results}\label{sec:Results}
In a Bayesian framework, the detection of a stochastic process is often approached through model selection. 
Previous studies have demonstrated that Galactic foreground can surpass the noise levels of space-borne detectors, potentially affecting the detection of other backgrounds~\cite{Bender:1997hs,Nelemans:2001hp,Barack:2004wc,Edlund:2005ye,Ruiter:2007xx,Nelemans:2009hy,Cornish:2017vip,Huang:2020rjf,Liang:2021bde,Staelens:2023xjn,Hofman:2024xar,Pozzoli:2024wfe}. 
Consequently, we focus on whether space-borne detectors can discern \acp{SGWB} of different origins amidst the Galactic foreground, rather than merely detecting \acp{SGWB} in the presence of pure detector noise. 

In this section, we will explore the following three hypotheses: (i) $\mathcal{H}_{0}$, which represents the sole presence of Galactic foreground; (ii) $\mathcal{H}_{1}$, involving Galactic foreground and one additional background; and (iii) $\mathcal{H}_{2}$, encompassing Galactic foreground and two additional backgrounds. 
After establishing that multi-component \acp{SGWB} can be detected and distinguished, the final step will be parameter estimation to determine the extent to which space-borne detectors can constrain the parameters of these backgrounds. 
It is worth emphasizing that, while the astrophysical background involves a variable index for $\Omega_{\rm ast}$, the cosmological background is assumed to have a fixed index of 0 for $\Omega_{\rm cos}$.

Unless otherwise stated, we generate simulated datasets following the methodology outlined in~Sec.~\ref{sec:method}, utilizing the cross-correlation of the four channel pairs $\{\rm AA', AE', EA', EE'\}$ within the TL network.

\subsection{Model selection}
We are currently engaged in calculating the log-Bayes factors across various \ac{SGWB} parameters, aiming to assess the detection capability of the TL network to \acp{SGWB} amidst the Galactic foreground. 
The intensity variation of the Galactic foreground over extended operation times is a key consideration in this assessment. 
Moving on to the astrophysical and cosmological backgrounds, we streamline the analysis by setting the astrophysical spectral index to be around 2/3 (specifically, within the range of [1/3, 1])\footnote{In this paper, we limit our focus to this simple case, rather than other complex ones, such as the background from extreme mass ratio inspirals~\cite{Bonetti:2020jku,Pozzoli:2023kxy,Piarulli:2024yhj}.}. 
This simplification contributes to the reasonableness of the forthcoming analysis, which will be elaborated upon later.

Given the flexibility of the model, we set the energy densities $\Omega_{\rm ast/cos}$ to be on the order of $10^{-12}$, and present the log-Bayes factor between the hypotheses $\mathcal{H}_{0}$ and $\mathcal{H}_{1}$ in~\fig{fig:B_ast_cos}. 
For an astrophysical background, the $\ln \mathcal{B}_{10}$ remains below 3 even after two years of cross-correlation detection. 
As the operation time extends to 4 years, an astrophysical with $\Omega_{\rm ast}$ of $7\times 10^{-12}$ can be detected, where the $\ln \mathcal{B}_{10}$ is larger than 5. 
Regarding a cosmological background, it can be detected with just 1 year of cross-correlation detection when the $\Omega_{\rm cos}$ reaches $9\times10^{-12}$, with a $\ln \mathcal{B}_{10}$ of around 5. 
When the operation time extends to 4 years, the intensity threshold of $\Omega_{\rm cos}$ can be further decreased to $5\times10^{-12}$, with a $\ln \mathcal{B}_{10}$ being greater than 5. 
\fig{fig:B_ast_cos} supports the notion that with extended operation time, the accumulation of cross-correlation data and the diminishing foreground intensity significantly boost the detection capability to other \acp{SGWB}.

The next step entails distinguishing between two backgrounds while the Galactic foreground is present. 
Detecting two backgrounds introduces new challenges compared to detecting a single one: the prior range of $\alpha_{\rm ast}$ for astrophysical background can encompass the value of 0. 
This scenario can result in a strong cosmological background being wrongly identified as an astrophysical background. 
Additionally, in the model selection process, the simpler of two models that fit the data equally well is preferred~\cite{Romano:2016dpx}. 
Consequently, a reduced Bayesian factor between hypotheses $\mathcal{H}_{1}$ and $\mathcal{H}_{2}$ can arise. 
To address this issue, we have previously set the astrophysical spectral index within the range of [1/3, 1]. 
Besides, we expand the energy density $\Omega_{\rm ast/cos}$ from $10^{-12}$ to $10^{-11}$ and set the operation time to 4 years.

Following the rules mentioned above, we employ the Galactic foreground corresponding to a 4-year operation time and set the $\Omega_{\rm ast}$ for astrophysical background at $2.64\times 10^{-11}$, which aligns with the median expected intensity of the extragalactic background. 
Furthermore, we increase the value of $\Omega_{\rm cos}$ from $1\times 10^{-11}$ to $5\times 10^{-11}$ to demonstrate the impact of the intensity of the cosmological background on the model selection. 
In~\fig{fig:B_tot}, we present the log-Bayes factor $\ln (\mathcal{B}_{21})$ between hypotheses $\mathcal{H}_{1}$ and $\mathcal{H}_{2}$ by the brown line, where the background in $\mathcal{H}_{1}$ specifically attributed to an astrophysical origin. 
For comparison purposes, the purple line denotes the log-Bayes factor $\ln (\mathcal{B}_{10})$ between hypotheses $\mathcal{H}_{0}$ and $\mathcal{H}_{1}$. 
With the intensities of backgrounds on the order of $10^{-11}$, the 4-year data of the TL network is sufficient to provide evidence for the existence of \ac{SGWB}, as $\ln (\mathcal{B}_{10}) \gg 5$. 
This result aligns with the results depicted in~\fig{fig:B_ast_cos}. 
Furthermore, for $\ln (\mathcal{B}_{10})>5$, $\Omega_{\rm cos}$ must exceed $4\times 10^{-11}$ to enable the TL network to identify the cosmological background among the Galactic foreground and astrophysical background.

\subsection{Parameter estimation}
Having established through model selection that the 4-year operation of the TL network is capable of detecting one or more \acp{SGWB} amidst the Galactic foreground, our subsequent focus shifts to parameter estimation. 
This process aims to determine the precision with which the TL network can constrain the parameters of these \acp{SGWB}. 
Notably, in contrast to model selection, the spectral index $\alpha_{\rm ast}$ now functions as a model parameter to be inferred.

We initiate the parameter estimation process for a scenario that encompasses the Galactic foreground and a single \ac{SGWB}. 
Adhering to the parameter settings employed during model selection, we establish $\Omega_{\rm ast}$ and $\Omega_{\rm cos}$ to $2.64\times10^{-11}$ and $1\times 10^{-11}$, respectively, within a 4-year Galactic foreground. 
The resulting posterior distributions are illustrated in~\fig{fig:cp_ast} and~\fig{fig:cp_cos} of the Appendix~\ref{appen:ORF}, where the corresponding P-P plot ensures the consistency of parameter estimation. 
Regarding the foreground parameter $a_{i}$, its posterior distribution widens with increasing order, indicating a growing uncertainty. 
The astrophysical \ac{SGWB} is delineated by two parameters, $\Omega_{\rm ast}$ and $\alpha_{\rm ast}$, which exhibit a strong correlation. 
This parameter correlation complicates the estimation process. 
In contrast, the cosmological \ac{SGWB}, characterized by the single parameter $\Omega_{\rm cos}$, allows for a more precise constraint that closely mirrors the true value. 
It is noteworthy that the parameters of foreground and \acp{SGWB} do not exhibit a strong correlation. 
This observation justifies the use of a polynomial fit for the foreground in \ac{SGWB} detection. 
To enhance clarity, we further consolidate the parameters of foreground and \acp{SGWB} into their spectral density parameter $\Omega_{\rm gw}$, with the posteriors displayed in the top panel of~\fig{fig:PE_tot}. 
Given that the Galactic foreground dominates near 1 mHz, its estimated median value closely matches the true value, albeit with a minor error range. 
Concerning the \acp{SGWB}, although the credible region is relatively expansive, the corresponding P-P plot can guarantee the consistency of parameter estimation.

In our ongoing efforts to estimate the parameters of multi-component \ac{SGWB}, we adjust the $\Omega_{\rm ast}$ of astrophysical background to the anticipated upper limit of $1.32\times10^{-10}$. 
Meanwhile, we increase the $\Omega_{\rm cos}$ of cosmological background to $5\times10^{-11}$~\cite{Liang:2021bde}. 
This adjustment allows us to examine the precision of our parameter estimation for the cosmological background in the presence of a foreground, particularly when the astrophysical background's intensity is at its theoretical maximum. 
We present the posterior distributions in~\fig{fig:cp_tot} of the Appendix~\ref{appen:ORF}, confirming that the true values of the corresponding ten parameters are indeed contained within the 1-$\sigma$ credible intervals. 
We observe that when multiple components are present in the \ac{SGWB}, correlations may emerge between the parameters of the different components. 
The resulting posteriors of energy spectrum density $\Omega_{\rm gw}$ are illustrated in the bottom panel of~\fig{fig:PE_tot}. 
When compared to the detection of a single \ac{SGWB}, the simultaneous detection of multiple \acp{SGWB} can lead to a loss in the precision of estimating the energy spectrum density. 
Nonetheless, under the model parameters chosen, the true values for both the astrophysical and cosmological backgrounds continue to lie within the 1-$\sigma$ credible regions. 
Our findings suggest that, following 4 years of operation of the TL network, there remains a promising prospect for detecting the cosmological background within permissible models, even in the presence of a foreground and an astrophysical \ac{SGWB}.

\section{Summary}\label{sec:Summary}
In this paper, we investigated the cross-correlation detection for the \ac{SGWB} using a space-borne detector network. 
We initiated our study by deriving a tailored likelihood function specifically for scenarios involving a strong \ac{SGWB}, which also enables data folding to accelerate processing. 
The likelihood is primarily concerned with the segment duration for data folding and the statistical error estimation of cross-correlation. Employing the TL network as a case study, we selected the segment duration to be 1 hour, taking into account the detection sensitivity band and the model error of the \ac{ORF}.
Meanwhile, adhering to this segmentation strategy, we employ the auto-correlation estimator to accurately estimate the statistical errors.

Building upon the established foundation, we utilized model selection and parameter estimation to forecast the detectable limits of the \ac{SGWB} in the presence of the Galactic foreground. 
To simplify our analysis, we have assumed a fixed zero-slope for the cosmological background. 
Our results revealed that, by extending the operation time of the TL network from 0.5 years to 4 years, the detectable limits to the \ac{SGWB} could be significantly improved. Specifically, for a single astrophysical or cosmological \ac{SGWB}, an intensity of $\Omega_{\rm ast/cos}$ on the order of $10^{-12}$ would be sufficient to furnish evidence of the \ac{SGWB}'s existence. 
In scenarios where both astrophysical and cosmological \ac{SGWB} coexist, the $\Omega_{\rm cos}$ should reach $2\times 10^{-11}$ to ensure the identification of the cosmological background.

For a single \ac{LISA}, Boileau et al. proposed that a cosmological background energy density of approximately $8\times 10^{-13}$ could be distinguished from a Galactic foreground and the anticipated astrophysical background~\cite{Boileau:2021sni}. 
In contrast to our work, their work has taken into account a variable index for the cosmological background. 
Nevertheless, this assessment did not consider the impact of detector noise uncertainties on the null-channel method. 
As detailed in~\cite{Muratore:2023gxh}, compared to scenarios without detector noise uncertainties, the background energy density would need to be roughly 50 times higher to achieve the same measurement precision in the presence of such uncertainties. 
Considering this factor, the cross-correlation detection of a detector network can outperform the null-channel detection of a single detector for the \ac{SGWB}.

We acknowledge the potential to expand upon our current work by delving into more complex detector noise characteristics, including non-stationary~\cite{Cornish:2014kda,Powell:2018csz,Robson:2018jly,Zackay:2019kkv,Davis:2022ird,Kumar:2022tto}, non-Gaussian~\cite{Allen:2001ay,Allen:2002jw,Himemoto:2006hw,Martellini:2015mfr,Yamamoto:2016bxj,Zackay:2019kkv}, and correlated noise~\cite{Thrane:2013npa,Thrane:2014yza,Coughlin:2016vor,Kowalska-Leszczynska:2016low,Himemoto:2017gnw,Meyers:2020qrb,Janssens:2021cta,Janssens:2021doe,Himemoto:2023keu,Janssens:2024jln}. 
Rather than hindering the SGWB detection if modeled properly, such complexities can potentially be used to enhance the discrimination between noise and signal. 
Additionally, the first order \ac{PT} with a broken power-law background can also emerge in the mHz band~\cite{Caprini:2009yp,Hindmarsh:2013xza,Jinno:2016vai,Jinno:2017fby,Hindmarsh:2017gnf,Cutting:2018tjt,Huang:2017laj,Mazumdar:2018dfl,Wang:2020jrd,Di:2020kbw,Wang:2021dwl,Jiang:2023nkj}. Through separating it from other backgrounds, on the one hand, the parameter estimation of remaining backgrounds can be more precise. On the other hand, the detection of the \ac{PT} background itself can also verify that the electroweak symmetry of elementary-particle physics is broken.

\begin{acknowledgments}
This work has been supported by the National Key Research and Development Program of China (No. 2023YFC2206704), the National Key Research and Development Program of China (No. 2020YFC2201400), the Natural Science Foundation of China (Grants No. 12173104), and the Guangdong Basic and Applied Basic Research Foundation(Grant No. 2023A1515030116). 
Z.C.L. is supported by the Guangdong Basic and Applied Basic Research Foundation (Grant No. 2023A1515111184). 
We also thank Jianwei Mei for the helpful discussions. 
\end{acknowledgments}

\clearpage

\appendix
\bw
\section{Results of the corner plot and P-P plot check}\label{appen:ORF}
For comprehensive presentation, we have included the corner plots for posterior distributions of the \ac{SGWB} parameters in this appendix, corresponding to the results presented in~\fig{fig:PE_tot}. Furthermore, we have conducted the corresponding P-P plots to guarantee consistency. For a completely consistent method, we anticipate that the P-P plot will follow the diagonal line with some random fluctuations. Note that, given the numerous parameters of the Galactic foreground, we present the P-P plot corresponding to the estimation of its intensity at 1 mHz.

\begin{figure}[b]
	\centering
	\includegraphics[width=.90\linewidth]{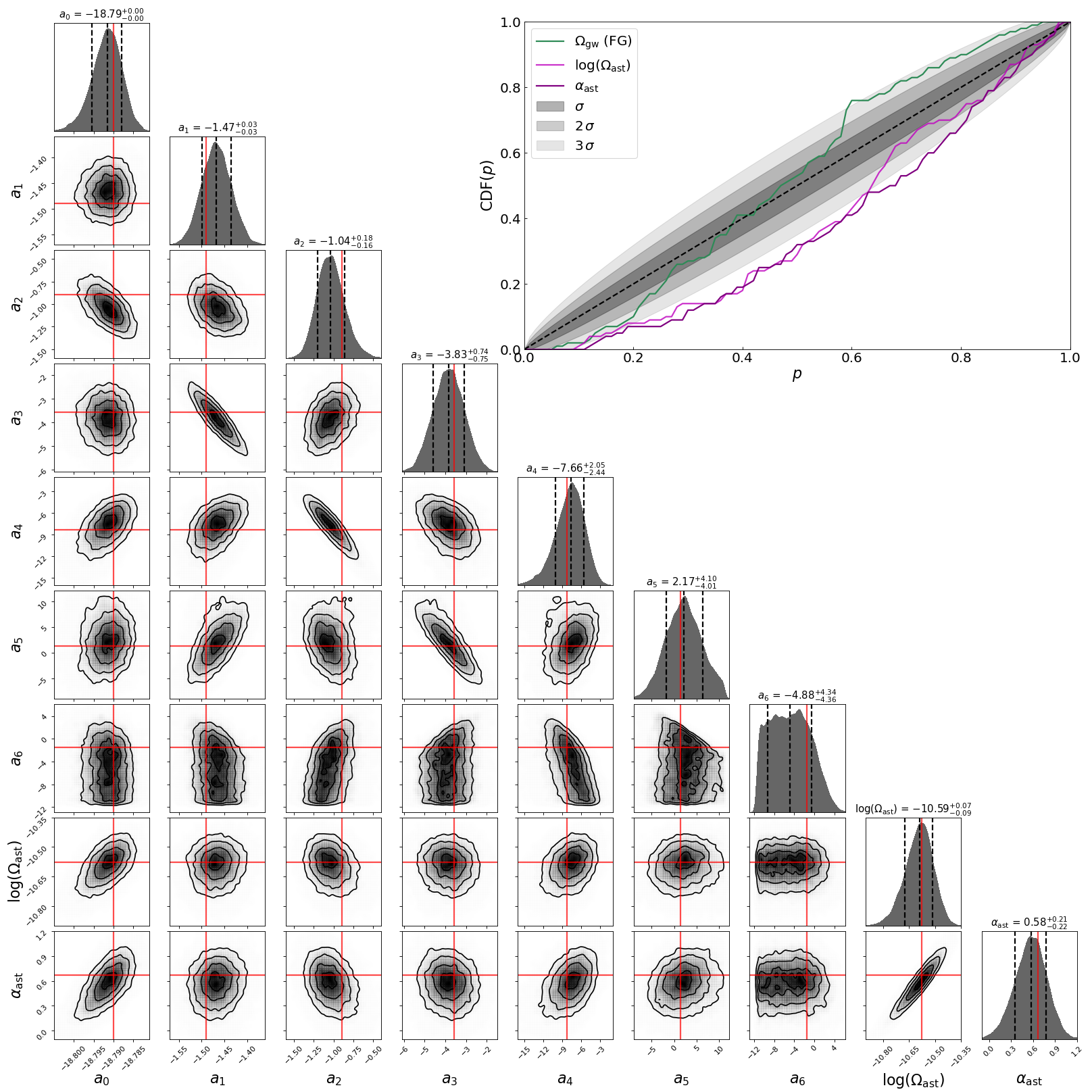}
	\caption{Corner plot depicting the posterior distributions, generated using  sampling. These results incorporate a 4-year Galactic foreground and an astrophysical background with $\Omega_{\rm ast}=2.64\times 10^{-11}$. Vertical dashed lines on the posterior distributions mark the quantiles [16\%, 50\%, 84\%], with red lines representing the true values. The inset presents the P-P plots from 100 independent simulations for estimating $\Omega_{\rm gw}$ (green) at 1 mHz of the Galactic foreground, $\log (\Omega_{\rm ast})$ (violet) and $\alpha_{\rm ast}$ (purple) of the astrophysical background. The $x$-axis represents the credible level, and the $y$-axis represents the cumulative distribution function (CDF), which is the proportion of simulations where the true parameters lie within the credible interval. Different gray-shaded areas represent the 1-$\sigma$, 2-$\sigma$, and 3-$\sigma$ confidence intervals, respectively.}
	\label{fig:cp_ast}
\end{figure}

\clearpage
\begin{figure}[h]
	\centering
	\includegraphics[width=.90\linewidth]{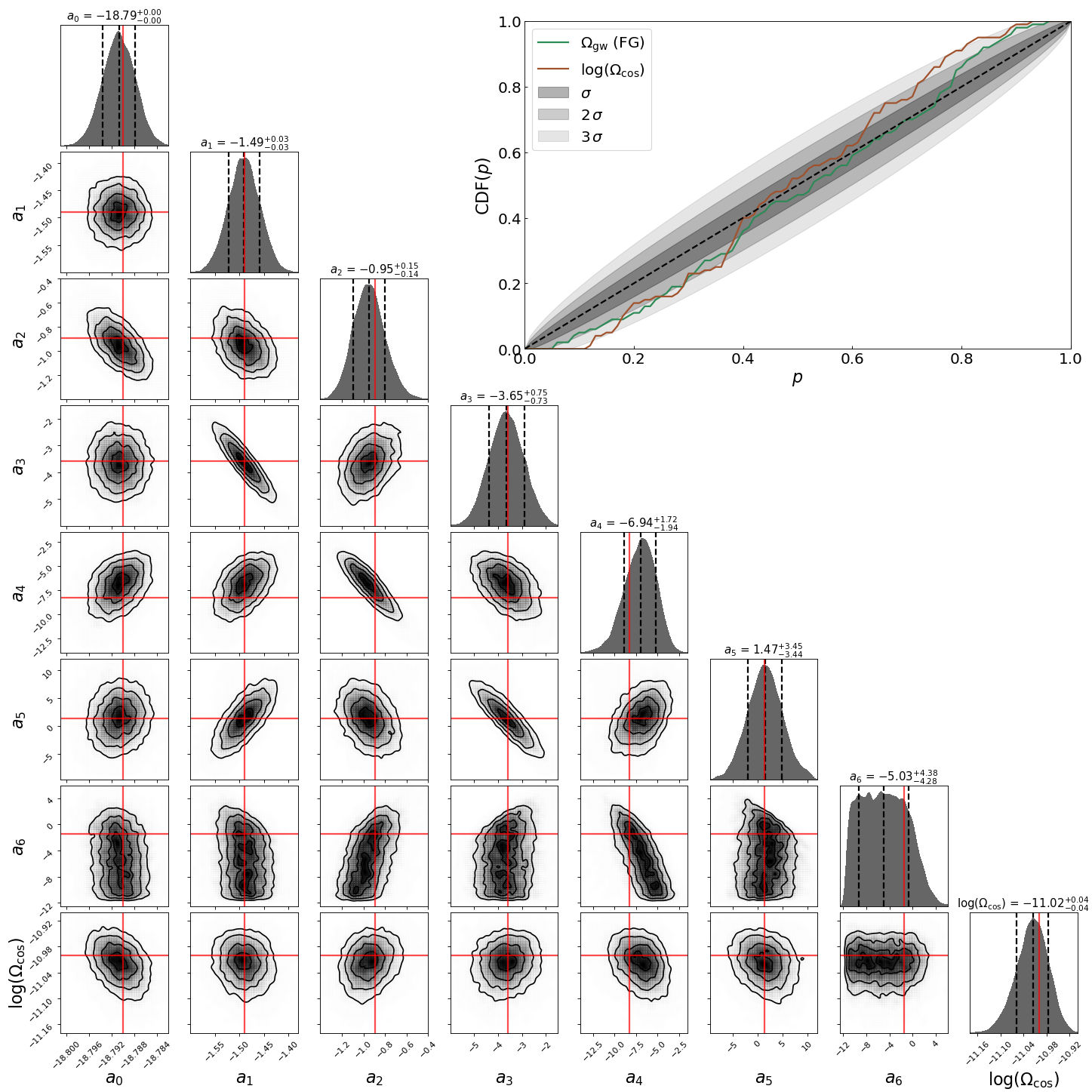}
	\caption{Corner plot depicting the posterior distributions, generated using sampling. These results incorporate a 4-year Galactic foreground and a cosmological background with $\Omega_{\rm cos}=1\times 10^{-11}$. Vertical dashed lines on the posterior distributions mark the quantiles [16\%, 50\%, 84\%], with red lines representing the true values. The inset presents the P-P plots from 100 independent simulations for estimating $\Omega_{\rm gw}$ (green) at 1 mHz of the Galactic foreground, $\log (\Omega_{\rm cos})$ (brown) of the cosmological background. The $x$-axis represents the credible level, and the $y$-axis represents the CDF, which is the proportion of simulations where the true parameters lie within the credible interval. Different gray-shaded areas represent the 1-$\sigma$, 2-$\sigma$, and 3-$\sigma$ confidence intervals, respectively.}
	\label{fig:cp_cos}
\end{figure}

\clearpage

\begin{figure}[t]
	\centering
	\includegraphics[width=.95\linewidth]{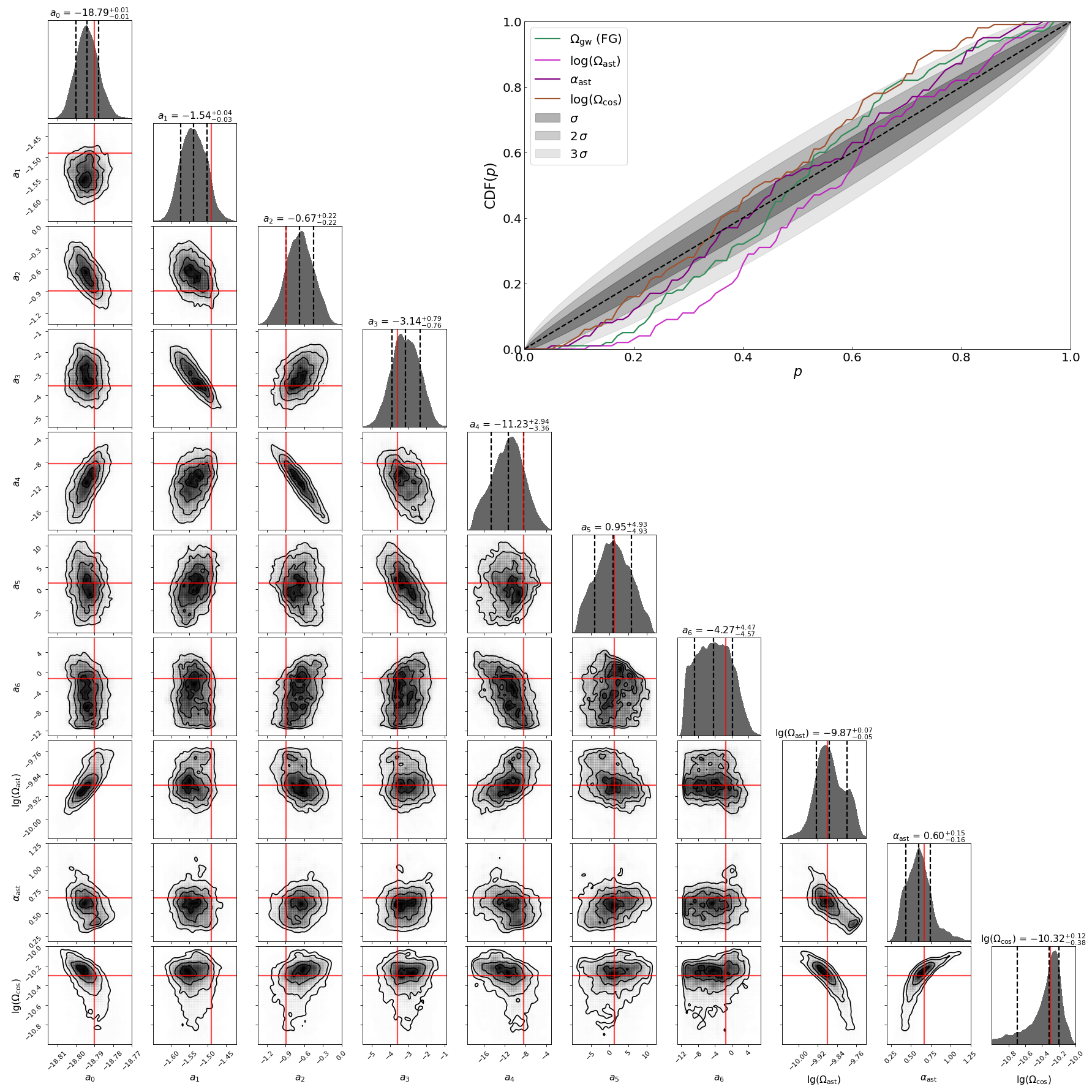}
	\caption{Corner plot depicting the posterior distributions, generated using sampling. These results incorporate a 4-year Galactic foreground, an astrophysical background with $\Omega_{\rm ast}=1.32\times 10^{-10}$, and a cosmological background with $\Omega_{\rm cos}=5\times 10^{-11}$. Vertical dashed lines on the posterior distributions mark the quantiles [16\%, 50\%, 84\%], with red lines representing the true values. The inset presents the P-P plots from 100 independent simulations for estimating $\Omega_{\rm gw}$ (green) at 1 mHz of the Galactic foreground, $\log (\Omega_{\rm ast})$ (violet) and $\alpha_{\rm ast}$ (purple) of the astrophysical background, $\log (\Omega_{\rm cos})$ (brown) of the cosmological background. The $x$-axis represents the credible level, and the $y$-axis represents the CDF, which is the proportion of simulations where the true parameters lie within the credible interval. Different gray-shaded areas represent the 1-$\sigma$, 2-$\sigma$, and 3-$\sigma$ confidence intervals, respectively.}
	\label{fig:cp_tot}
\end{figure}
\ew


\normalem
\bibliographystyle{apsrev4-1}
\bibliography{ref.bib}

\end{document}